\documentstyle[11pt,newpasp,twoside]{article}
\def\edcomment#1{\iffalse\marginpar{\raggedright\sl#1\/}\else\relax\fi}
\reversemarginpar

\begin{document}
\title{Perhaps they are not globular clusters after all}
 \author{A. M. Karick}
\affil{School of Physics, University of Melbourne, Victoria 3010, Australia}

\begin{abstract}
Our 2dF Fornax Cluster Spectroscopic Survey (FCSS) and follow-up work
in the Virgo Cluster have shown that the cores of both galaxy clusters
contain a previously-unknown class of object, ultra-compact dwarf (UCD) galaxies. We
present high resolution spectroscopy and deep multicolour imaging to
show that these enigmatic objects are dynamically distinct from both
globular clusters (GCs) and nucleated dwarf galaxies (dE,Ns). Our
hypothesis for their origin may explain the observed high ``specific
frequency'' of GCs in central cluster galaxies.
\end{abstract}

\section{Dynamical Analysis and Stellar Populations}
Internal velocity dispersions of the UCDs and a comparison dE,N
(FCC303) have been measured using the VLT and Keck Telescopes
(Drinkwater et al.2003). The velocity dispersions of the UCDs range
from 24 to 37 km s$^{-1}$, considerably higher than those of Galactic
GCs. The UCDs lie well off the globular cluster L $\propto$
$\sigma$$^{1.7}$ relation in a previously unoccupied region. This
supports the ``galaxy threshing'' model in which the UCDs are the
remnant nuclei of infalling dE,Ns which have been tidally disrupted by
the cluster cD galaxy (Bekki et al. 2001). Removing the halo of the
dE,N galaxy reduces the total luminosity by about a factor of 100 but
barely changes the central velocity dispersion. We obtained deep
multicolour imaging (u,g,r,i) of the core of the Fornax Cluster taken
with the CTIO MOSAIC camera. The colours of the UCDs were compared
with a sample of NGC 1399 globulars (Mieske et al. 2003) and the
nuclei of cluster dE,Ns. Preliminary analysis suggests that the UCDs
have colours similar to the nuclei of the brightest dE,Ns and GCs. \\

\noindent {\bf Acknowledgements}\\ This work was done in collaboration
with M. J. Drinkwater, M. D. Gregg and  M. Hilker.

\end{document}